\documentclass[amsmath,a4paper,twocolumn,pre,floatfix,showpacs]{revtex4-1}

\usepackage{braket}
\usepackage{graphicx}
\usepackage{xcolor}
\usepackage{dcolumn}
\usepackage{bm}
\usepackage{amssymb}

\begin{document}

\title{Superradiance at the localization-delocalization crossover in tubular chlorosomes}

\author{Rafael A. Molina}

\author{Enrique Benito-Mat\'{\i}as}

\affiliation{Instituto de Estructura de la Materia, IEM-CSIC, Serrano 123, Madrid 28006, Spain}

\author{Alejandro Somoza}

\author{Lipeng Chen}

\author{Yang Zhao}

\affiliation{Division of Materials Science, Nanyang Technological University, 50 Nanyang Avenue, Singapore 639798}

\begin{abstract}
We study the effect of disorder on spectral properties of tubular chlorosomes
in green sulfur bacteria {\em Cf.~aurantiacus}. Employing a Frenkel-exciton Hamiltonian with diagonal and off-diagonal disorder consistent
with spectral and structural studies, we analyze excitonic localization and spectral statistics of
the chlorosomes. A size-dependent localization-delocalization crossover is found to occur as a function of the excitonic energy. The crossover energy region coincides with the more optically active states with maximized superradiance, and is, consequently, more conducive for energy transfer.
\end{abstract}


\maketitle


\section{Introduction}
\label{sec:intro}
Chlorosomes are the main light-harvesting structures of green sulfur bacteria, some green filamentous
anoxygenic phototrophs \cite{Frigaard06}, and the recently discovered aerobic anoxygenic phototroph {\em Candidatus
Chloroacidobacterium thermophilum} \cite{Bryant07}. The chlorosomes are self-assembled structures of hundreds of thousands bacteriochlorophyll (BChl) molecules. They differ from other light-harvesting complexes by the {absence} of a
protein matrix {which} supports the {photosynthetic} pigments and their very large size with lengths up to 200 nm \cite{Holzwarth90}.
Chlorosomes usually function in extremely low light conditions and are thus probably the most efficient light-harvesting
antenna complexes in nature \cite{Blankenship_book}.
These special properties {makes the chlorosome} a potential candidate for use in biomimetic light-harvesting devices \cite{Modesto10,Sridharan09}.
Due to the heterogeneity of chlorosomes their structure cannot be determined by crystallographic methods and structural information is scarce. However, thanks to recent advances in different areas, a structural model has been put forward for cholorosomes of the green sulfur bacterium {\em Chlorobaculum tepidum} \cite{Ganapathy09}. This model is based
on a {\em syn-anti} array of BChl c pigments in tubular shape. One of the methods that were used to determine
the structure of the chlorosome was to study mutant bacteria with three genes whose expression was prevented. The mutant bacteria have a more ordered structure of BChls which allowed X-ray crystallographic studies and showed the presence of an helical structure. Interestingly enough, the mutant bacteria growth was much slower than the wild type bacteria. This observation together with the fact that the function of three genes was suppressed in order to make the structure less disordered hints to the possibility that disorder plays a biological function in chlorosomes and
a disordered structure is favored by evolution. The obvious role for disorder is the broadening of the optical spectra that in an ordered structure would be too narrow as observed in similar artificial supramolecular structures \cite{Ganapathy09}.

However, these results are paradoxical since disorder is also expected to localize the excitonic wave functions and inhibit energy diffusion according to the theory of Anderson localization. Since the seminal paper of Anderson on the absence of diffusion for disordered quantum lattices \cite{Anderson58}, wave localization due to interference has been a central theme in condensed matter physics. One of the main tools
for studying localization phenomena and, in general, disordered quantum systems has been Random Matrix Theory (RMT)
\cite{Mehta_Book}. The RMT approach has also been applied to probe thermodynamic properties of closed systems as well as transport properties of open systems. One of the important results in the field is the statistical relationship between the spectral repulsion
parameter $\beta$ and the wave function localization. The $\beta$ parameter can be used to locate the metal-insulator transtion in three dimensions \cite{Evers08} and it has been shown that there are specific scaling laws between localization and the repulsion parameter in finite, disordered one-dimensional systems \cite{Izrailev89,Izrailev90,Casati93,Sorathia12}. Localization in systems with dipolar interactions like the natural light-harvesting complexes have been shown to be much weaker than for the Anderson model with only nearest neighbor coupling \cite{Rodriguez03}. Quantum transport in disordered networks has been extensively studied in connection with its relevance for efficient energy transfer in light-harvesting systems. Centrosymmetry and a dominant doublet spectral structure has been stablished as a general mechanism for highly efficient quantum transport even in the presence of disorder and the natural implementation of this mechanism in the Fenna-Matthews-Olson (FMO) { complex} of green sulfur 
bacteria has also been explored \cite{Walschaers13}.  

The extraordinary energy transfer properties of chlorosomes are attributed to cooperative phenomena in the BChl aggregates, which are known otherwise as superradiance and supertransfer. Superradiance occurs when a group of emitters, situated at a distance from each other shorter than the wave length of the light, emits coherently with a very high intensity, and was discovered by Dicke in the context of radiating atoms \cite{Dicke54}. It has
been studied for supramolecular complexes and in the light-harvesting apparatus of bacterial photosynthesis \cite{Meier97,Zhao99}.
In the presence of disorder, static or dynamic,
superradiance is quenched, although, 
in a tight-binding model in a ring superradiance persists for disorder strengths below a critical value even in a regime with strong coupling to the external environment {\cite{Celardo14}}.  The optical properties of
one-dimensional disordered excitonic systems have been studied extensively, specially in the context of applications to J-aggregates {\cite{Knapp84,aggregates_book}}. The localization length of the system has been shown to limit the number of monomers contributing to the superradiance {\cite{Fidder91}}. These results should also apply to
supertransfer, a coined term describing non-radiative excitonic transfer of a superradiant patch {\cite{Strek77,Lloyd2010, Abasto12}}. {Supertransfer has been ascribed to play a crucial role in the efficiency enhancement of light-harvesting systems, and in the chlorosome in particular \cite{Kassal2013, Huh2014}. Photosynthetic systems are arranged in hierarchical structures where energy funnelling from the chlorosome antenna towards the reaction center proceeds throughout a network of superradiant units. Supertransfer refers to the excitation energy transfer (EET) between these superradiant patches}. The properties of cylindrical chlorosomes have been studied before in the homogeneous case \cite{Didraga02} and in the disordered case \cite{Didraga04}. Attempts have also been made at describing the excitonic dynamics in the presence of a bath using a high-temperature, stochastic treatment attributed to Haken and Strobl \cite{Fujita12,Yejun12,Moix13}.

In this paper, we study the effects of disorder on the localization properties, the spectral statistics and the optical properties, in particular, superradiance, of chlorosomes. We try to give an answer to the previously
mentioned paradox. The long range properties of the dipolar interaction between chromophores make the excitonic
wave functions more robust against localization. { However}, for realistic values of disorder and size, there appears a localization-delocalization crossover as a function of the excitonic energy as monitored by the inverse participation ratio and the spectral statistics. Precisely in the low-energy region where
there { appears} more collectiveness in the optical behavior of the states for the clean system, {the exciton is} more localized when disorder is included. { In the presence of static disorder, there} is {thus a} compromise between {the collective but localized }low energy {states} ({with }less nodes in the wave function) and delocalization which happens at
high energy. { We find that the} maximum of superradiance appears at the localization-delocalization crossover
and moves to the high energy side of the spectrum as disorder increases. An optimal value of disorder depending on the environmental conditions exists between sufficient superradiance and a wide spectrum that could help explain the energy transfer properties
of chlorosomes. The plan of the paper is the following: we present the realistic model for a single rod in Sec. \ref{sec:model}, the results for the spectral statistics, wave function localization and optical properties are shown in Sec. \ref{sec:results} and, finally, some conclusions are drawn in Sec. \ref{sec:conclusions} with some comments on the possibility of probing experimentally these theoretical results.


\section{Model for a single rod}
\label{sec:model}

\begin{figure}
\includegraphics[width=\linewidth]{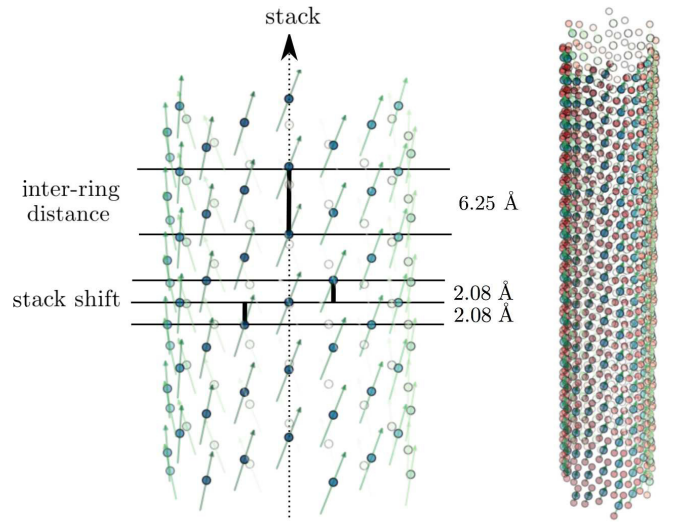}
\caption{\label{fig:structure} Sketch of the arrangement of pigments in the chlorosome complex of {\em Cf. aurantiacus}. The structure is { formed} by a series of $L$ concentric rings separated by 6.25 \AA. A stack is defined as a vertical group of sites, which share the same dipole orientation. The structure has { 18 stacks} so the total number of sites in the full structure is {$N=18L$}. Additionally, there is a vertical shift of 2.08 {\AA~} in between adjacent stacks inducing helicoidal pathways in the structure. Each BChl c presents a transition dipole moment (green arrows) which approximately connects two nitrogens (red dots) situated diagonally to the magnesium atom (blue dots) of the chromophore at the center of the pigment.}
\end{figure}

We make use of a single rod aggregate structure consisting of a series of $L$ concentric rings in conformance with the model introduced in \cite{Prokhorenko00,Holzdata}, as shown in Fig.~\ref{fig:structure}. The distance between consecutive rings is 6.25 \AA. The structure has {18} different stacks. A stack is defined as a vertical group of sites sharing the same dipole orientation, { or equally, the number of sites in one ring}. Each ring {contains therefore} {18 sites} and the total number of molecules of the full structure is {$N=18L$}. For a certain pigment, the neighbouring sites in the same ring belonging to neighboring stacks present relative heights of -2.08 {\AA} and 2.08 {\AA} with respect {to the} cylinder symmetry axis. Every pigment in the same stack { has} the same dipole orientation, forming an angle of 100$^\circ$ with the radius vector connecting the magnesium atom and the symmetry axis and an angle of 36.5$^\circ$ with this axis. Regarding the dipole moments of the pigments belonging to the same ring, neighboring dipole moments are rotated with each other 20$^\circ$ 
along the symmetry axis. { The radius of the cylinder is 21.13 \AA~ (the distance from the magnesium atoms to the symmetry axis).} This structure results
in helical pathways of the excitons for energy transfer from the center of the structure to {the} top or bottom \cite{Ganapathy09,Prokhorenko03,Oostergetel10}.

The excitonic model we use to treat optical properties and localization is based on a model of Frenkel excitons delocalized over a molecular lattice.
\begin{equation}
 H=\sum_{n,m} J_{nm} a^{\dagger}_n a^{\phantom{\dagger}}_m,
 \label{eq:Frenkel}
\end{equation}
where $J_{nm}$ refers to the excitonic coupling among different sites via dipole-dipole interaction, being
$J_{nn} = \epsilon_n$ the monomer excitation energy of a single site. The operators $a^{\dagger}_n$ ($a^{\phantom{\dagger}}_n$) create (destroy)
a molecular excitation at site $n$. { As is often the case within the literature of photosynthetic excitons, we will remain in the one-exciton manifold, assuming that at every time there is only one exciton in the system.} Due to the high density of pigments within the chlorosome complex, the often used point dipole approximation is not valid in the present regime where the dipole moment lengths are comparable to the distance between {different} chromophores. A better approximation is given by the extended dipole-dipole interaction \cite{Czikklely70}. A point dipole is substituted by opposite charges $+\delta$ and $-\delta$ at a finite distance $l$:

\begin{equation}
 J_{nm}=\frac{D^2}{4 \pi \epsilon \epsilon_0 l^2} \left( \frac{1}{r_{++}} + \frac{1}{r_{--}}-\frac{1}{r_{+-}}-\frac{1}{r_{-+}} \right).
\end{equation}

\noindent where $r_{\pm \pm}$ represent the distance between the positive (negative) charge of the point dipole associated with the first molecule to
the positive (negative) charge of the point dipole associated with the second molecule. In the case of BChl c pigments the dipole length $l$ is taken to be 8 \AA~ and the
squared dipole strength $D$ as 25 $\mathrm{Debye}^2$.
The values for the disorder are taken from Ref.~\cite{Prokhorenko03} and are quoted on Table \ref{table:dis}.
For understanding how disordered the system really {is} we can compare {these values to} the parameters of the system. The value for the nearest-neighbor coupling in the system {(between adjacent pigments in the same stack)} is
{$-282.39$} cm$^{-1}$, the FWHM of the diagonal disorder is 210 cm$^{-1}$ while the FWHM of the nearest-neighbor coupling induced by the non-diagonal disorder is 162.22 cm$^{-1}$. In the case of the Anderson model with only nearest-neighbor coupling {we have verified that} the system is fully localized and superradiance is quenched for these values of disorder and for the realistic typical sizes of the chlorosomes. The situation is quite different for the full dipole-dipole interaction as is presented in the next section. {For the nearest-neighbors model we have retained the coupling between  adjacent sites in the same stack and ring, being the latter $~10$ times smaller and of opposite sign than the former, due to their relative dipole orientation.}

\begin{table}
\begin{tabular}{cc}
 \hline
 Disorder & value of FWHM \\
 \hline
 Exciton diagonal energy  & 210 cm$^{-1}$ \\
 Position along symmetry axis &  0.24 \AA \\
 Position along x-axis        &  0.3  \AA \\
 Position along y-axis        &  0.3  \AA \\
 angle between dipole and symmetry axis &  10$^{\circ}$ \\
 angle between dipole and radius-vector &  20.6$^{\circ}$ \\
\end{tabular}
 \caption{\label{table:dis} Parameters for the disorder used in our work, taken from ~\cite{Prokhorenko03}.}

\end{table}

\section{Results: Localization, spectral statistics and optical properties}
\label{sec:results}

In the absence of disorder, an infinite, pristine system can be diagonalized analytically in the momentum representation, and eigenstates can be labeled with
two momenta, the longitudinal one on the symmetry axis of the cylinder with continuous values and the transverse
one with discrete values that depend on the number of molecules in each ring. Following Didraga {\em et al.} \cite{Didraga02}, it is possible to calculate the absorption spectra from the Bloch states that diagonalize the Hamiltonian
in momentum state. There are two peaks that concentrate all the {oscillator} strength. Most of the oscillator {strength} is concentrated at the excitonic lower state at energy $E_0=\omega_0+\Sigma'_n J(n)$, where $\omega_0$ is the energy of one exciton of an isolated BChl c molecule and $\Sigma'_n J(n)$ represents the sum of all the couplings in site space. The wave function of
this state is just { a} coherent superposition of all { dipoles} with the same coefficient, it has zero transverse momentum {$k_{\bot}=0$}. The other peak is related to the helical structure of the system, the {corresponding} states have transverse momentum {$k_{\bot}=\pm 1$}. The amplitude in this state is such that the components of the {dipoles} parallel to the {z-axis} interfere destructively while the {ones} perpendicular to the z-axis add constructively (the opposite as in the other case). In the case of a finite system without disorder the helical peak is more fragmented due to the effect of the ending points of the cylinder. Although, the monomers are
placed in a three-dimensional cylindrical structure wrapped in itself, the properties of the model are more similar to the properties of a quasi one-dimensional system with dipolar interactions \cite{Didraga02,Didraga04}. It is interesting to notice that the clean model {of the} chlorosome has a centrosymmetric structure that has been argued to be an important ingredient for efficient transport in {quantum} networks {\cite{Christandl05,Kay06,Zech13}}. A dominant doublet structure is the other critical ingredient needed for optimal quantum  {transport \cite{Walschaers13}}. Although, outside the model we are using in this work, it is easy to picture a
dominant doublet structure as being also important for the behavior of chlorosomes once the full modelling of the coupling to the reaction center is taking into account.

Didraga and Knoester also studied a case of strong disorder with
with $\sigma=800$ {cm$^{-1}$} in addition to a weak-disorder case where disorder can be seen
as a perturbation mixing helical states with the low energy peak and fragmenting slightly the {oscillator} strength.
They showed that the excitons dominating the optical properties have anisotropic localization along the helices and
that the localization length increased with energy from the bottom to the center of the band. In their case, however
the localization length was always smaller {than} the typical system size.

\begin{figure}
 \includegraphics[width=0.45\textwidth]{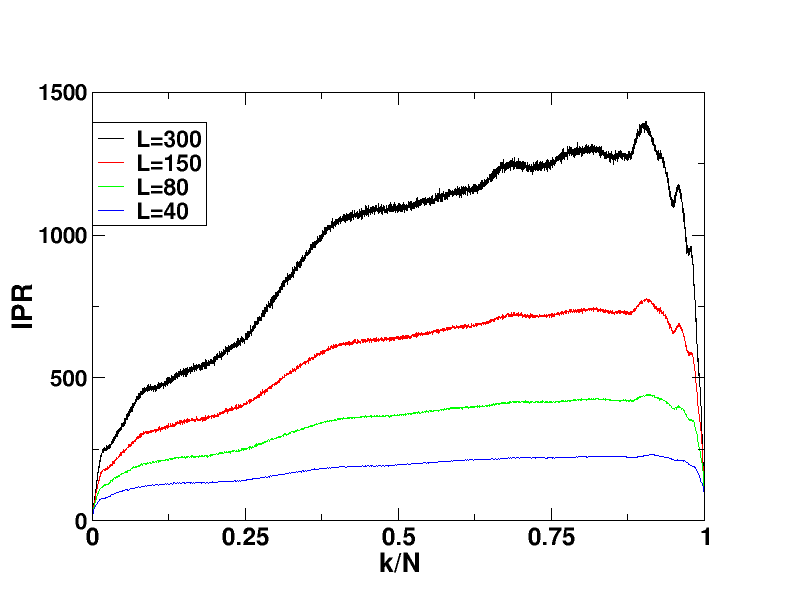}
 \includegraphics[width=0.45\textwidth]{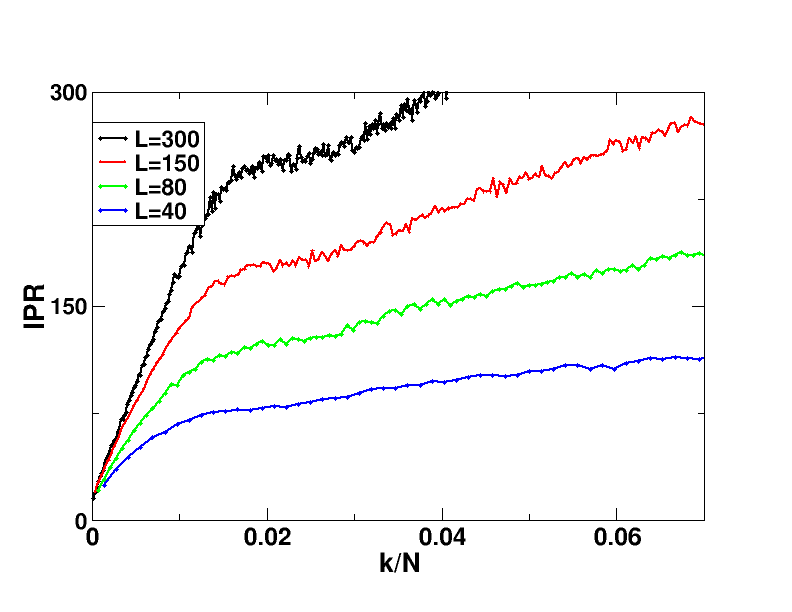}
 \caption{\label{fig:ipr} IPR as a function of the normalized state number $k/N$ for different sizes. The lower panel shows a zoom on the region close to the ground state.
 We can see how IPR saturates as a function of size for the region close to the ground state up to $k/N \approx 0.005$. Parameters for the disorder are quoted in Table \ref{table:dis}.
 The FWHM of the diagonal disorder is 210 cm$^{-1}$ while the FWHM of the nearest-neighbor coupling induced by the non-diagonal disorder is 162.22 cm$^{-1}$.}
\end{figure}

In order to study the localization properties of the excitonic system we calculate the inverse participation ratio (IPR) \cite{Bell70} defined by
\begin{equation}
 {\rm IPR}(E_{\alpha})=\frac{1}{\sum_{i} |\Psi_i^{\alpha}|^4},
\end{equation}
where $\Psi_i^{\alpha}$ is the amplitude of the excitonic eigenstate with energy $E_{\alpha}$ on site $i$. ${\rm IPR}=1$ if the state is localized on only one site, and ${\rm IPR}=N$ if the wave function is equally distributed among all sites in an $N$ site system.
Fig.~\ref{fig:ipr} displays the ${\rm IPR}$ as a function of the normalized state number $k/N$ for cylinders of different lengths. We notice that for realistic values of disorder the IPR is independent of the aggregate length at the edges of the spectrum  and thus, wavefunctions present a localized nature. Therefore, the IPR reveals  a localized to delocalized crossover which is also signalled by the spectrum statistics as shown below.

The repulsion parameter is defined by the behaviour of the nearest neighbour spacing distribution near degeneracy: $P(s)\sim s^\beta \; (s\rightarrow0)$, where $s_{\alpha}=(E_{\alpha +1}-E_{\alpha})$. Integrable systems present a Poissonian profile of the spacing {distribution} with $\beta=0$ while chaotic systems show level repulsion, avoiding degeneracies with $\beta \sim 1$. The origin of level repulsion comes from the presence of avoided crossings in systems with symmetry-breaking perturbations. Recently, level repulsion has been experimentally verified in disordered cyanine-dye-based molecular nanoaggregates \cite{Augulis10}.
In order to study the repulsion parameter $\beta$ of spectral statistics, we fit the nearest neighbor spacing distribution to the phenomenological Brody distribution \cite{Brody73} which has been used succesfully for similar purposes in many areas related to localization in disordered systems or quantum chaos \cite{Wilkinson91,Molina01,Flores13,Frisch14,Mur15}:

\begin{equation}
 P(s)= A(\beta+1)s^{\beta} \exp({-As^{\beta+1}}). \,\,A=\Gamma\left[\frac{\beta+2}{\beta+1}\right]^{\beta+1}
\end{equation}
where the constant $A$ is needed for proper normalization {and $\Gamma$ is the Gamma function}. For localized systems, level repulsion is diminished as neighboring levels tend to localize in non-overlapping spatial regions, $\beta=0$ and the nearest neighbor spacing distribution is equal to the Poisson distribution $P(s)=\exp{(-s)}$. In systems of disordered but extended states the form of $P(s)$ can be obtained from RMT and can be approximated by the Wigner-Dyson distribution which is equal to the Brody distribution with $\beta=1$, $P(s)=(\pi/2)\exp{(-\pi s^2/4)}$ \cite{Mehta_Book}. The Brody distribution interpolates smoothly between these two limits.
Previously to the computation of the $P(s)$, the unfolding of the spectra to unit spacing is needed. The unfolding procedure filters the smooth part of the spectrum by performing a local average on the nearest neighbour distances. We have used a local unfolding
procedure with a window of $5$ levels as described in \cite{Haake_book}. This procedure describes perfectly well the short
range spectral correlations described by the $P(s)$ although it cannot be used to study long range correlations in the spectra \cite{Gomez02}.

\begin{figure}
 \includegraphics[width=0.45\textwidth]{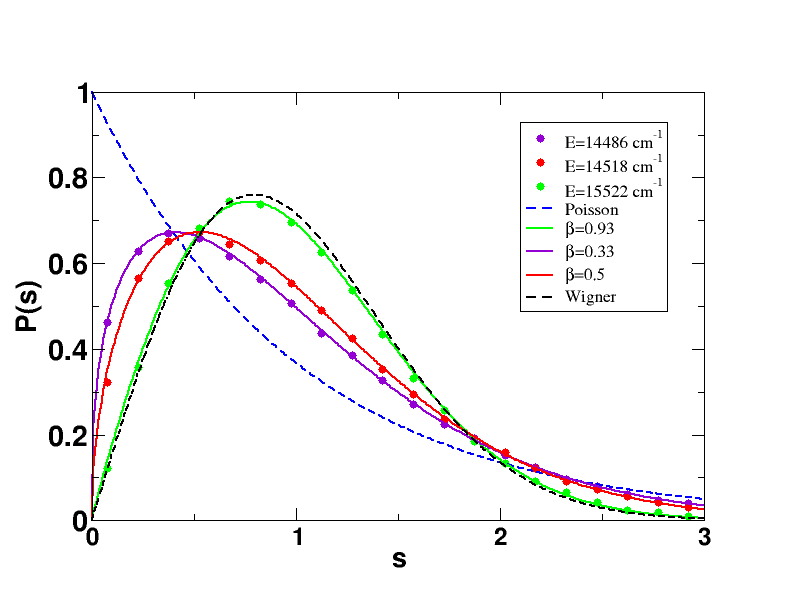}
 \includegraphics[width=0.45\textwidth]{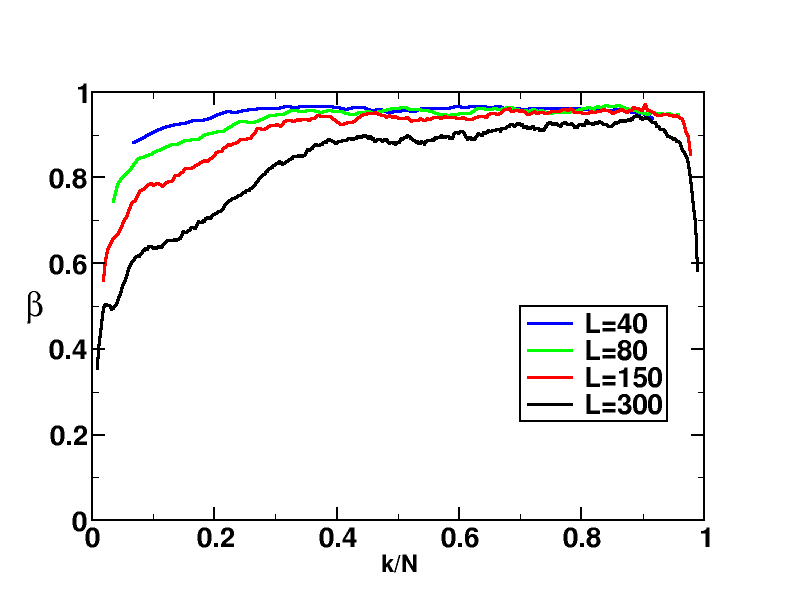}
 \caption{\label{fig:brody} Top panel: Fit of the $P(s)$ to the Brody distribution in the case of $L=40,80,150,300$ rings for different values
 of the central energy of the level window. Bottom panel: Value of $\beta$ as a function of the renormalized state number $k/N$ for
 different lengths of cylinder. Parameters for the disorder are the same as in Fig. \ref{fig:ipr} and are quoted in Table \ref{table:dis}.
 The FWHM of the diagonal disorder is 210 cm$^{-1}$ while the FWHM of the nearest-neighbor coupling induced by the non-diagonal disorder is 162.22 cm$^{-1}$.}
\end{figure}

The results are shown in Fig.~\ref{fig:brody}. We show some examples of the fit to the Brody distribution. We have analyzed the data moving a window of $100$ levels around some particular level number and analyzing the results every $10$ levels (the windows are overlapping). In that way, we can study the repulsion parameter as a function of the exciton energy. The quality of the fit is impressive. 
The parameter $\beta$ behaves as a function of the energy in a complementary way to the ${\rm IPR}$. {This complementarity is tightly related to the sensitivity of the $\beta$ parameter to the ratio between the localization length and the total length of the system \cite{Casati93,Sorathia12}}. {In the localized region, the IPR saturates for sufficiently large tubes, indicative of the finite localization length of these states. At higher energies, far from the ground state, the localization length is much longer than the realistic system sizes that we have used (300 rings). For this reason the IPR of Figure \ref{fig:ipr}  increases with larger aggregates in the delocalized region. }
 In contrast, the repulsion parameter decreases as a function of the system size in the localized region, as in that regime $\beta$ is very sensitive to the ratio between the localization length {(which saturates)} and the total length of the system. In the delocalized region, $\beta$ varies very little with the system size although it {continues} to decrease, albeit slowly. The fact that the repulsion parameter decreases (although very slightly in the middle of the band) as a function of the total length of the chlorosome for all energies is an indication of the localization of the wave functions in the thermodynamic limit,  as expected from theoretical considerations \cite{Rodriguez03,Evers08}.
{ In other words, while the localization length increases with the system size, the ratio localization length / system size tends to zero.} For the largest aggregate size we have calculated, the values we find numerically are $\beta \gtrsim 0.6$ in the delocalized region, $\beta \approx 0.5$ in the crossover region and $\beta \lesssim 0.4$ in the localized region of the spectra. { The value $\beta\approx 0.5$ corresponds to the energy region where the saturation of the IPR with the system size ceases, giving rise to extended states as shown in the bottom panel of Figure \ref{fig:ipr}}. Due to the finite nature of the system, the crossover region is slightly extended around this value.  

\begin{figure}
 \includegraphics[width=0.5\textwidth]{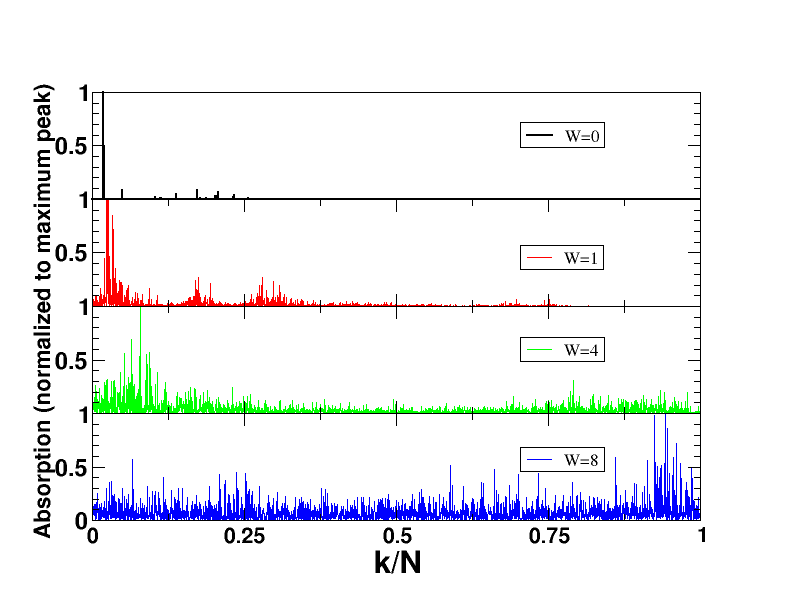}
 \caption{\label{fig:abs} Stick spectrum for the absorption (normalized to the maximum peak in each case) for four different values of disorder $W$, a factor multiplying the disorder parameters of Table \ref{table:dis} (From top to bottom, $W=0$, $W=1$, $W=4$, and $W=8$). The maximum of the spectrum is moving right
 with increasing disorder. In the case of $W=8$ superradiance has been destroyed and the {oscillator} strength is distributed along all the spectrum, in this particular realization the maximum happens to be by chance closer to the top end of the spectrum.
 The FWHM of the diagonal disorder is 0 cm$^{-1}$ ($W=0$), 210 cm$^{-1}$ ($W=1$), 840 cm$^{-1}$ ($W=4$), and 1680 cm$^{-1}$ ($W=8$) while the FWHM of the nearest-neighbor coupling induced by the non-diagonal disorder is 162 cm$^{-1}$ ($W=1$), 272 cm$^{-1}$ ($W=4$), and 359 cm$^{-1}$ ($W=8$). Note that the parameter $W$ multiplies the values of the FWHM for the geometric parameters of the model which does not translate directly in the same scaling of the FWHM of the nearest-neighbor coupling.}
\end{figure}

In general, the excitonic superradiance depends on a variety of controlling parameters, such as temperature, inter-chromophore coupling, static disorder, and various forms of exciton-phonon interactions. For example, the superradiance size is quickly reduced to one with rising temperature. It was also shown that the effect of exciton-phonon coupling on the superradiance size closely resembles that of the static disorder \cite{Meier97}.
Here, considering only static disorder, the superradiance size is limited by two factors: 1) it cannot be larger than the localization length; and 2)
it cannot be larger than the typical size of the nodal structure (the inverse of the absolute value of the momentum for systems with translational symmetry). In the pristine system when the states are delocalized over the entire system, the localization length is
not a limiting factor, and most of the {oscillator} strength is concentrated in the state with zero momentum. In the disordered
case as the localization length increases with energy while the nodal size decreases with energy there will be a maximum of {superradiance} when the localization length is approximately equal to the nodal size and there is a state
with coherence along the whole localization region. As disorder is increased the region with maximum superradiant
size moves to higher values of energy up to the point where disorder is so large {that} superradiance is destroyed and the oscillator strength becomes uniform in energy. In order to study superradiance we {concentrate} on the linear absorption spectrum that can be computed in the dipolar approximation through:
{
\begin{equation}
 A(\omega)=\sum_k d_k  \delta(\omega-E_k) ,
 \label{eq:absorption}
\end{equation}
\begin{equation}
d_k=\left| \sum_n \braket{n|\psi_k} \boldsymbol{\mu}_{n}  \right|^2=\sum_{n,m} (c^{k}_n)^* c^k_m  ~\boldsymbol{\mu}_{n}  \cdot \boldsymbol{\mu}_{m}
 \label{eq:osc}
\end{equation}
where $d_k$ is the oscillator strength of excitonic state $k$}, $\psi_k$ is the $k$ excitonic wave function, $E_k$ is the corresponding energy, { $\ket{n}$ represents a local excitation at site $n$, $c_n^k$ is the $n$-th component of the $k$ excitonic wave function} and $\boldsymbol{\mu}_{n}$ is the dipole vector corresponding to the $n$ molecular dipole in the structure. We assume an isotropic distribution of chlorosomes and average over the direction of the polarization vector.
We show some examples of the linear absorption stick spectrum as a function of the level number for typical realizations with different disorder strengths
in Fig. \ref{fig:abs}.

\begin{figure}
 \includegraphics[width=0.45\textwidth]{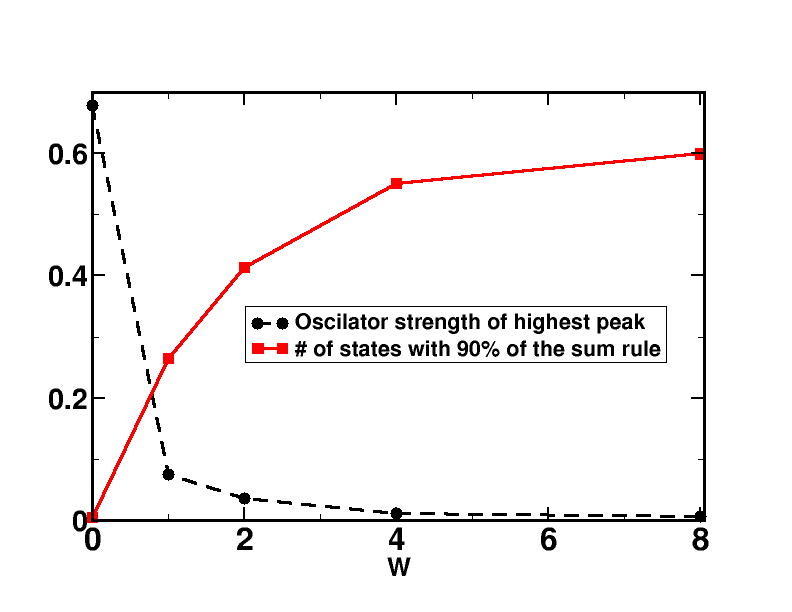}
 \caption{\label{fig:superradiance} {Oscillator} strength of the highest peak in the stick absorption spectrum in units of the sum rule and
 number of states with a combined {oscillator} strength of $90\%$ in units of the total number of states as a function
 of disorder (defined as in the previous figure) for the case of $L=150$ rings. The result shown is the ensemble average with $1000$ realizations of
 the disorder.}
\end{figure}

We have examined how {absorption} spectra of the system and spectral statistics are related to superradiance and localization
of the wave function.
As a measure of the superradiance length as a function of disorder, we have calculated the oscillator strength of
the highest peak. In order to measure the fragmentation, we have also calculated the minimum number of states that have a combined {oscillator} strength equal to $90\%$ of the sum rule (we first order the states in augmenting oscillator strength and sum up to the $90\%$ of the sum rule).
In Fig. \ref{fig:superradiance} we show both these measures as a function of the strength of disorder $W$ in units of
the parameters of table \ref{table:dis}. The FWHM of diagonal disorder is then $210~ W$ cm$^{-1}$ while the behavior of the non-diagonal disorder is more complicated as the FWHM of the geometric parameters of the model does not translate directly into a FWHM of the non-diagonal couplings which depend on this geometry (see caption of Fig. \ref{fig:abs}). We can see that superradiance is important up to a disorder strength of
$W \sim 4$ corresponding to a FWHM of the diagonal disorder of 840 cm$^{-1}$ and to a FWHM of the nearest-neighbor coupling of 272 cm$^{-1}$. Although we have examined only a single rod model for numerical convenience, we expect the results to hold for realistic cholorosomes with many concentric rods.

\section{Conclusions}
\label{sec:conclusions}

Superradiance and supertransfer are two sides of the same coin and key to understand the fast energy transfer in chlorosomes of green sulfur bacteria. We have studied the effect of disorder in superradiance of chlorosomes of {\em Cf.~aurantiacus} using the excitonic model developed recently by Ganapathy {\em {et} al.} \cite{Ganapathy09}. The dipolar long-range interaction makes disorder not very effective for localizing the excitonic states for cylindrical chlorosomes. However, using tools of Anderson localization theory and spectral statistics such as the IPR and the repulsion parameter $\beta$ we have shown that there is a localization-delocalization crossover as a function of the excitonic energy that is seen both in the wave function statistics and in the spectral statistics. This kind of supramolecular aggregates concentrate the optical active energy region at low energies were the collective effects of the coupled molecular dipoles are most important. There is, then, a
competition between localization and collectiveness (or state coherence) and the more optical active region where superradiance occurs coincides with the crossover region between localization and delocalization. The results found in this work should be applicable to other different supramolecular aggregates \cite{Zhao00,Hu07}. These states where the absorption is largest are also the ones expected to dominate the energy transfer. The collective supertransfer effect is still important in spite of disorder and should increase dramatically the single dipole transfer rates. There has been experiments that have shown level repulsion via analysis of
the fluoresence spectra in J-aggregates \cite{Malyshev07,Augulis10}. Similar experiments as a function of the size could be performed for chlorosomes to probe the level repulsion
in the optically active region and as a consequence the localization of the wave functions with respect to the total size of the chloromeses. The results should be different in the wild type case, more disordered, than for less disordered mutants. 

It would also be very interesting to extend the results obtained to the case of coupling to a thermal environment. Studying the transport and absorption properties of the one-dimensional Anderson model coupled to a bath, it has been shown that the maximal diffusion rate occurs at intermediate coupling strength \cite{Moix13}. These results are reminiscent of our results as a function of the energy. These theoretical observations have been linked to the phenomenom of {\em momentum rejuventation} apart from the role of dephasing in the destruction of destructive interference leading to Anderson localization \cite{Li15}. The momentum rejuvenation occurs when classical noise counteracts the depletion of high momentum components of the wave-packet suistaining a broad momentum distribution. Realistic models, such as the one explored in this work, in the presence of a thermal bath should be explored in order to understand better the role of these general design principles \cite{Walschaers13} in the actual behavior 
of chlorosomes as light-harvesting complexes.    


This work was supported in part by Spanish MINECO project FIS2012-34479, CSIC project I-Link0938, CAM research consortium QUITEMAD+ S2013/ICE-2801, and the Singapore National Research Foundation through the Competitive Research Programme under Project No.~NRF-CRP5-2009-04.

\end{document}